# The anomalous electron magnetic moment and Lamb shift in the new approach in quantum electrodynamics

V. A. Golovko

Moscow Polytechnic University

Bolshaya Semenovskaya 38, Moscow 107023, Russia

E-mail: fizika.mgvmi@mail.ru

Abstract

In a previous paper by the author a new approach in quantum electrodynamics was proposed in which the electronic and electromagnetic fields are ordinary *c*-numbers in contradistinction to noncommuting *q*-numbers used in the standard formulation of quantum electrodynamics. The factual development of standard quantum electrodynamics began when the anomalous correction to the electron magnetic moment and the Lamb shift were calculated and the calculations were confirmed by experiment. It is demonstrated in the present paper how the anomalous electron magnetic moment and Lamb shift can be obtained in the new approach in quantum electrodynamics.

**1. Introduction**

In Ref. [1] (hereafter referred to as I) a new approach in quantum electrodynamics (NAQED) was proposed in which the electronic and electromagnetic fields are ordinary *c*-numbers in contradistinction to noncommuting *q*-numbers used in standard quantum electrodynamics (SQED). The electronic wave function ψ in NAQED represents a real distribution of the electronic density in space (equal to |ψ|$^2$) instead of the usual probabilistic interpretation of the wave function [2]. In this connection the question arises as to why a free electron remains localized in space despite the electric repulsion between its parts charged identically. The existence of the spin of the electron indicates that the electron rotates and thereby there are electric currents in the interior of the electron whereas the currents flowing in one direction attract each other. The magnetic attraction alone is, however, insufficient. Consequently, an amount of charge of opposite sign is required. That is the physical reason why the electron in NAQED is described by two bispinors ψ$_1$ and ψ$_2$ where the second bispinor ψ$_2$ corresponds to a charge of opposite sign. The equilibrium of all forces is possible only when the total charge *e* of the electron has a strictly definite value, which amounts to saying that the dimensionless fine-structure constant $\alpha = e^2/\hbar c$ has a strictly definite value as well. Upon solving the equations of NAQED, which can be done only numerically, one will simultaneously find the value of α. All of these are considered in detail in I. It may be added that the necessity of the rotation for the appearance of the attractive magnetic forces signifies that the electron exists only owing to its rotation characterized by its spin ½ℏ while the rotation and charge of the electron are interrelated, which is embodied in the rigorous value of $\alpha = e^2/\hbar c$.

SQED in spite of its tremendous success is incapable of determining the structure of the electron and thereby the value of α because its wave function ψ is a *q*-number whereas in experiment one can observe only quantities which are *c*-numbers. It should be stressed that nowhere in SQED does one consider the electron to be point-like. A point-like particle is represented by a delta function. Nowhere in SQED does one represent the electron by the delta function. For example, when one considers scattering of a photon by the electron, one represents the electron by a density matrix which is not the delta function, see § 86 of [3]. In integrals corresponding to Feynman diagrams, the electron is not represented by the delta function either. Could the electron be described with the help of the delta function, this would greatly simplify involved calculations in SQED. Although SQED indicates that the electron is not point-like, its methods give no way of determining the electron structure.

According to I a free electron has the size lying in the range between the electron Compton wavelength and the Bohr radius in a hydrogen atom. It is explained in I that not only does such a large size of the electron not contradict experiment, but on the contrary it is supported by all experimental data concerning quantum mechanics.

Successful development of SQED began in the late 1940s when methods permitting one to calculate the anomalous correction to the electron magnetic moment and the Lamb shift were work out. In the present paper we shall demonstrate how the anomalous electron magnetic moment and Lamb shift can be obtained in NAQED.

In Sec. 2 of the paper, we write down basic equations required in the following sections. Section 3 is devoted to calculating the normal electron magnetic moment according to NAQED while in Sec. 4 we calculate the anomalous electron magnetic moment. In Sec. 5 we turn our attention to the Lamb shift. The results obtained are discussed in the concluding section.

Throughout the present paper, if the opposite is not pointed out, we utilize the definitions and notation adopted in [3] which coincide with the ones of I. It should be underlined that all quantities and functions in this paper are *c*-numbers.

**2. Basic equations**

In the present study we shall restrict ourselves to first in $\alpha$ corrections to different quantities. It follows from Eq. (4.3) of I that $\psi_2 \sim \alpha\psi_1$, which is natural since $\psi_2$ is a secondary bispinor with respect to the primary bispinor $\psi_1$. We now see from Eq. (4.2) of I that the terms with $\psi_2$ are in magnitude of the order $\alpha^2\psi_1$. Thus, in our study we can neglect all herms with $\psi_2$.

According to the end of Sec. 3 of I, if there is an external electromagnetic field, the four-potential $A_\mu$ is to be replaced by $A_\mu + A_\mu^{\text{ext}}$ in the equations of motion where $A_\mu^{\text{ext}}$ is the external four-potential while the equation for the electromagnetic field tensor remains unchanged. Having this in mind we write out Eq. (5.1) of I upon neglecting $\psi_2$ and writing $\psi$ instead of $\psi_1$:

$$ic\hbar\gamma^\mu \frac{\partial \psi}{\partial x^\mu} - e\left(A_\mu + A_\mu^{\text{ext}}\right)\gamma^\mu\psi - \frac{c\hbar}{\lambda}\psi = 0. \qquad (2.1)$$

We recast Eq. (3.8) of I analogously:

$$\frac{\partial F^{\mu\nu}}{\partial x^\nu} = -4\pi e j^\mu, \ \ j^\mu = \overline{\psi}\gamma^\mu\psi, \qquad (2.2)$$

where $F^{\mu\nu} = \partial A^\nu/\partial x_\mu - \partial A^\mu/\partial x_\nu$ is the electromagnetic field tensor and $j^\mu$ is the particle current density. Equations (2.1) and (2.2) constitute a complete set of equations for $\psi$ and $A_\mu$.

It is helpful to find the Lagrangian which yields the above equations. The Lagrangian density in the present case can readily be obtained from the strict Lagrangian density of (3.3) of I:

$$L(\mathbf{r},t) = -\frac{1}{16\pi} F_{\mu\nu} F^{\mu\nu} + \frac{ic\hbar}{2}\left(\overline{\psi}\gamma^\mu \frac{\partial \psi}{\partial x^\mu} - \frac{\partial \overline{\psi}}{\partial x^\mu}\gamma^\mu \psi\right) - e\left(A_\mu + A_\mu^{\text{ext}}\right)\overline{\psi}\gamma^\mu \psi - \frac{c\hbar}{\lambdabar}\overline{\psi}\psi. \quad (2.3)$$

It is worthy of remark that the above equations as well as the strict equations of I including both the bispinors $\psi_1$ and $\psi_2$ do not contain the electron mass $m$. This mass makes its appearance in NAQED in the non-relativistic limit alone [1]. The absence of the mass in the theory opens up possibilities of calculating the mass spectrum of leptons. The leptons apart from the electron correspond to exited states of the electronic field. Upon calculating the energy $E$ of such states one will find the mass of the relevant leptons by the relation $E = mc^2$ valid in NAQED in the non-relativistic limit. If one computes the lifetime of the exited states, one will find the lifetime of the relevant leptons.

It is convenient to work with dimensionless quantities, denoting them by a tilde, as in I:

$$t = \frac{\lambdabar}{c}\tilde{t},\ x^k = \lambdabar \tilde{x}^k,\ \psi = \frac{1}{\lambdabar^{3/2}}\tilde{\psi},\ A_\mu = \frac{e}{\lambdabar}\tilde{A}_\mu,\ F^{\mu\nu} = \frac{e}{\lambdabar^2}\tilde{F}^{\mu\nu},\ L = \frac{c\hbar}{\lambdabar^4}\tilde{L}. \quad (2.4)$$

In what follows we shall omit the tilde bearing in mind that, if an equation does not contain $\hbar$, $c$, $e$ or $m$ but contains $\alpha$, the equation is written in the dimensionless units.

Equations (2.1) and (2.2) become

$$i\gamma^\mu \frac{\partial \psi}{\partial x^\mu} - \alpha\left(A_\mu + A_\mu^{\text{ext}}\right)\gamma^\mu \psi - \psi = 0, \quad (2.5)$$

$$\frac{\partial F^{\mu\nu}}{\partial x^\nu} = -4\pi j^\mu,\ j^\mu = \overline{\psi}\gamma^\mu \psi, \quad (2.6)$$

while the Lagrangian density of (2.3) will be

$$L(\mathbf{r},t) = -\frac{\alpha}{16\pi} F_{\mu\nu} F^{\mu\nu} + \frac{i}{2}\left(\overline{\psi}\gamma^\mu \frac{\partial \psi}{\partial x^\mu} - \frac{\partial \overline{\psi}}{\partial x^\mu}\gamma^\mu \psi\right) - \alpha\left(A_\mu + A_\mu^{\text{ext}}\right)\overline{\psi}\gamma^\mu \psi - \overline{\psi}\psi. \quad (2.7)$$

### 3. The normal electron magnetic moment

According to Eq. (4.21) of I the magnetic moment $\boldsymbol{\mu}$ is given by

$$\boldsymbol{\mu} = e\lambdabar\tilde{\boldsymbol{\mu}} = \frac{e\hbar}{mc}\tilde{\boldsymbol{\mu}},\quad \tilde{\boldsymbol{\mu}} = \tfrac{1}{2}\int_{(\infty)} [\mathbf{r}\mathbf{j}]dV. \quad (3.1)$$

The second expression for $\boldsymbol{\mu}$ is written in units accepted usually for the magnetic moment where $m$ is the electron mass seeing that $\lambdabar = \hbar/(mc)$. The quantity $\tilde{\boldsymbol{\mu}}$ is dimensionless.

If we choose the cylindrical coordinates as in I, we shall see that $\tilde{\boldsymbol{\mu}}$ has only one component $\tilde{\mu}_z$ which is given by Eq. (4.22) of I. In that equation we neglect the functions $f_\nu^{(2)}$ which are relevant to $\psi_2$ and omit the superscript (1) in accord with (2.5), so that

$$\tilde{\mu}_z = \int\limits_{(\infty)} \rho \left(f_1 f_4 - f_2 f_3\right) dV . \tag{3.2}$$

Equation (2.5) with $A_\mu^{\text{ext}} = 0$ yields Eqs. (4.6) – (4.9) of I which, after substituting Eq. (4.26) of I, in the present case acquire the form

$$\frac{\partial f_4}{\partial \rho} + \frac{\partial f_3}{\partial z} + \frac{f_4}{\rho} - \alpha A f_4 + \alpha(c_0 + \overline{\varphi}) f_1 = 0, \tag{3.3}$$

$$\frac{\partial f_3}{\partial \rho} - \frac{\partial f_4}{\partial z} + \alpha A f_3 + \alpha(c_0 + \overline{\varphi}) f_2 = 0, \tag{3.4}$$

$$\frac{\partial f_2}{\partial \rho} + \frac{\partial f_1}{\partial z} + \frac{f_2}{\rho} - \alpha A f_2 + (2 - \alpha c_0 - \alpha \overline{\varphi}) f_3 = 0, \tag{3.5}$$

$$\frac{\partial f_1}{\partial \rho} - \frac{\partial f_2}{\partial z} + \alpha A f_1 + (2 - \alpha c_0 - \alpha \overline{\varphi}) f_4 = 0, \tag{3.6}$$

We have from (3.5) and (3.6) that

$$f_3(\rho,z) = -\frac{1}{2-\alpha c_0 - \alpha \overline{\varphi}} \left( \frac{\partial f_2}{\partial \rho} + \frac{\partial f_1}{\partial z} + \frac{f_2}{\rho} - \alpha A f_2 \right), \tag{3.7}$$

$$f_4(\rho,z) = -\frac{1}{2-\alpha c_0 - \alpha \overline{\varphi}} \left( \frac{\partial f_1}{\partial \rho} - \frac{\partial f_2}{\partial z} + \alpha A f_1 \right). \tag{3.8}$$

Now the combination that figures in (3.2) is

$$f_1 f_4 - f_2 f_3 = -\frac{1}{2-\alpha c_0 - \alpha \overline{\varphi}} \left[ \frac{1}{2} \frac{\partial}{\partial \rho} \left(f_1^2 - f_2^2\right) - \frac{\partial}{\partial z} f_1 f_2 - \frac{f_2^2}{\rho} + \alpha A \left(f_1^2 + f_2^2\right) \right]. \tag{3.9}$$

If we place this in (3.2), we shall get

$$\tilde{\mu}_z = -\int_0^{2\pi} d\varphi \int_{-\infty}^{\infty} dz \int_0^{\infty} \frac{\rho^2 d\rho}{2-\alpha c_0 - \alpha \overline{\varphi}} \left[ \frac{1}{2} \frac{\partial}{\partial \rho} \left(f_1^2 - f_2^2\right) - \frac{\partial}{\partial z} f_1 f_2 - \frac{f_2^2}{\rho} + \alpha A \left(f_1^2 + f_2^2\right) \right]. \tag{3.10}$$

The integrand does not contain the angle $\varphi$ and thereby the integration over $\varphi$ gives $2\pi$. All functions that figure in (3.10) vanish at infinity. If functions $F$ and $G$ vanish at infinity, one has

$$\int_0^{\infty} G\rho^2 \frac{\partial F}{\partial \rho} d\rho = -2\int_0^{\infty} \rho G F d\rho - \int_0^{\infty} F\rho^2 \frac{\partial G}{\partial \rho} d\rho, \quad \int_{-\infty}^{\infty} G \frac{\partial F}{\partial z} dz = -\int_{-\infty}^{\infty} F \frac{\partial G}{\partial z} dz . \tag{3.11}$$

If $G = 1/(2-\alpha c_0 - \alpha \overline{\varphi})$, then

$$G' = \frac{\alpha \overline{\varphi}'}{(2-\alpha c_0 - \alpha \overline{\varphi})^2}, \tag{3.12}$$

where the prime denotes differentiation with respect to $\rho$ or $z$.

If the integral in (3.10) is transformed with use made of (3.11) and (3.12) and all terms proportional to $\alpha$ are disregarded, the result will be

$$\tilde{\mu}_z = \pi \int_{-\infty}^{\infty} dz \int_0^{\infty} d\rho \rho f_1^2 . \tag{3.13}$$

The functions $f_\nu$ are subject to the normalization condition of Eq. (4.18) of I which in the present notation is of the form

$$\int_{-\infty}^{\infty} dz \int_0^{\infty} d\rho \rho \left( f_1^2 + f_2^2 + f_3^2 + f_4^2 \right) = \frac{1}{2\pi}. \tag{3.14}$$

It is seen from (3.3) that $f_3 \sim f_4 \sim \alpha f_1$. The function $f_2$ satisfies Eq. (A.1) of I which in the present case acquires the form

$$\nabla^2 f_2 - \frac{f_2}{\rho^2} - \nu^2 f_2 + \rho z Q_2 = 0, \tag{3.15}$$

where

$$Q_2 = -\frac{\partial G_3^{(1)}}{\rho \partial \rho} + \frac{\partial G_4^{(1)}}{z \partial z} - (2 - \alpha c_0) G_2^{(1)}, \tag{3.16}$$

whereas the quantities that enter in (3.16) are given in Eqs. (4.56)–(4.58) of I, namely,

$$G_2^{(1)} = \alpha \overline{\varphi} g_2^{(1)} + \alpha \gamma g_3^{(1)}, \quad G_3^{(1)} = \alpha \overline{\varphi} g_3^{(1)} + \alpha \rho^2 \gamma g_2^{(1)}, \quad G_4^{(1)} = \alpha \overline{\varphi} g_4^{(1)} - \alpha \gamma g_1^{(1)}. \tag{3.17}$$

The functions that figure in (3.17) are defined in Eq. (4.24) of I:

$$f_1 = g_1^{(1)}(\rho^2, z^2), \, f_2 = \rho z g_2^{(1)}(\rho^2, z^2), \, f_3 = z g_3^{(1)}(\rho^2, z^2), \, f_4 = \rho g_4^{(1)}(\rho^2, z^2), \, \gamma = A/\rho, \tag{3.18}$$

These equations together with (3.16) shows that $Q_2 \sim \alpha$ and thereby $f_2 \sim \alpha f_1$. Recalling that $f_3 \sim f_4 \sim \alpha f_1$ we see now that the terms containing $f_2$, $f_3$ and $f_4$ in (3.14) can be neglected in the limit as $\alpha \to 0$, in which case Eq. (3.13) yields $\tilde{\mu}_z = \frac{1}{2}$. As a result, it is seen from (3.1) that the normal electron magnetic moment is

$$\mu = \frac{e\hbar}{2mc} . \tag{3.19}$$

We have in NAQED obtained the well-known result. At the same time it should be underlined that $\mu$ of (3.19) is the normal magnetic moment of a free electron whereas in SQED one calculates the magnetic moment of the electron placed in a magnetic field. We postpone the discussion of these facts until the concluding section.

There is another way of calculating the magnetic moment. We shall in the next section employ this way in calculating the anomalous electron magnetic moment. As a preliminary step we shall elucidate in this section how this way can be used in computing the normal electron magnetic moment.

We revert to the Lagrangian density of (2.7). Any small correction to the Hamiltonian density amounts to a change in the sign of the Lagrangian density [3, § 42]. Therefore the

7correction to the Hamiltonian itself due the external potential $A_\mu^{\text{ext}}$ is the potential energy $\Pi$ whose form is seen from (2.7):

$$\Pi = \alpha \int_{(\infty)} \overline{\psi}\gamma^\mu \psi A_\mu^{\text{ext}} dV . \tag{3.20}$$

We further follow Bogoliubov and Shirkov [4, Sec. 30.2]. We represent the bispinor $\psi$ in terms of a plane wave

$$\psi = \frac{1}{\sqrt{2\varepsilon}} u(p) e^{-i\varepsilon t + i\mathbf{pr}} . \tag{3.21}$$

where $\varepsilon^2 = 1 + \mathbf{p}^2$. When considering the bispinor amplitude $u(p)$ we follow Ref. [3, § 23] where $u_p = u(p)$. It should be kept in mind, however, that we imply the dimensionless units of (2.4), so that in all formulae of [3] one should put $m = 1$ for the electron mass. In particular,

$$\overline{u}(p)u(p) = 2, \quad \overline{u}(p)\gamma^\mu u(p) = 2p^\mu , \tag{3.22}$$

where $p^0 = \varepsilon$. We represent the bispinor $u(p)$ in terms of two two-components quantities $\eta$ and $\chi$ (to avoid confusion with the potential $\varphi$ we use $\eta$ instead of $\varphi$ of [3]) as

$$u(p) = \begin{pmatrix} \eta \\ \chi \end{pmatrix}, \quad \chi = \frac{\mathbf{p}\sigma}{\varepsilon + 1} \eta , \tag{3.23}$$

where $\sigma$ is the Pauli vector matrix. The quantity $\eta$ can be written as

$$\eta = \sqrt{\varepsilon + 1}\, w , \tag{3.24}$$

in which $w$ is an arbitrary two-component quantity with the components $w_1$ and $w_1$ subject only to the normalization

$$w_1^* w_1 + w_2^* w_2 = 1 . \tag{3.25}$$

In what follows we shall decompose some functions into Fourier integrals according to

$$F(\mathbf{r}) = \frac{1}{(2\pi)^3} \int_{(\infty)} F(\mathbf{k}) e^{i\mathbf{kr}} d\mathbf{k}, \qquad F(\mathbf{k}) = \int_{(\infty)} F(\mathbf{r}) e^{-i\mathbf{kr}} dV . \tag{3.26}$$

In this context, it will be noted that the delta function is given by

$$\delta(\mathbf{k}) = \frac{1}{(2\pi)^3} \int_{(\infty)} e^{i\mathbf{kr}} dV . \tag{3.27}$$

We substitute the bispinors of (3.21) with slightly different $\varepsilon'$, $\mathbf{p}'$ and $\varepsilon$, $\mathbf{p}$ into (3.20), but we put $\varepsilon' = \varepsilon$ from the outset in order not to clutter up formulae. We transform $A_\mu^{\text{ext}}$ into a Fourier integral by (3.26). With use made of (3.27) we obtain

$$\Pi = \frac{\alpha}{2\varepsilon} \overline{u}(\mathbf{p}')\gamma^\mu u(\mathbf{p}) A_\mu^{\text{ext}}(\mathbf{k}), \qquad \mathbf{k} = \mathbf{p}' - \mathbf{p} . \tag{3.28}$$

We presume that the external scalar potential $\varphi \equiv A_0 = 0$, so that $\gamma^\mu A_\mu^{ext}(\mathbf{k}) = \gamma^0 A_0 - \gamma \mathbf{A}^{ext}(\mathbf{k}) = -\gamma^0 \boldsymbol{\alpha} \mathbf{A}^{ext}(\mathbf{k})$ where $\boldsymbol{\alpha}$ is the well-known vector matrix [3]. Recalling that $\bar{u} = u^* \gamma^0$ one has

$$\Pi = -\frac{\alpha}{2\varepsilon} u^*(\mathbf{p}') \boldsymbol{\alpha} \mathbf{A}^{ext}(\mathbf{k}) u(\mathbf{p}). \tag{3.29}$$

We imply that the external field varies only slowly in space, in which case $\mathbf{k}$ is small (in the following we shall take the limit $\mathbf{k} \to 0$).

We substitute (3.23) and take into account Eq. (33.5) of [3], namely,

$$(\boldsymbol{\sigma}\mathbf{a})(\boldsymbol{\sigma}\mathbf{b}) = \mathbf{ab} + i[\mathbf{ab}]. \tag{3.30}$$

As a result we have

$$\Pi = -\frac{\alpha}{2\varepsilon(\varepsilon+1)} \eta^*(\mathbf{p}')\{i\boldsymbol{\sigma}[\mathbf{kA}^{ext}] + (\mathbf{p}+\mathbf{p}')\mathbf{A}^{ext}\}\eta(\mathbf{p}). \tag{3.31}$$

The intensity of the magnetic field is

$$\mathbf{H}(\mathbf{r}) = \text{curl}\mathbf{A}(\mathbf{r}) = \frac{i}{(2\pi)^3}\int_{(\infty)}[\mathbf{kA}]e^{i\mathbf{kr}}d\mathbf{k}. \tag{3.32}$$

Therefore the Fourier transform of $\mathbf{H}(\mathbf{r})$ is $i[\mathbf{kA}]$. The last term in (3.31) is not relevant to our problem whereas the first term yields

$$\Pi = -\frac{\alpha}{2}w^*(\mathbf{p})\boldsymbol{\sigma}\mathbf{H}^{ext}w(\mathbf{p}), \tag{3.33}$$

in which (3.24) is used. We have also put $\mathbf{p}' = \mathbf{p}$ and $\varepsilon = 1$ in the non-relativistic limit. The coefficient of $\mathbf{H}^{ext}$ in (3.33) is the magnetic moment, so that

$$\boldsymbol{\mu} = \frac{e\hbar}{2mc}w^*(\mathbf{p})\boldsymbol{\sigma}w(\mathbf{p}) \tag{3.34}$$

in ordinary units according to (3.1). With use made of (3.25) we see that, if $|w_1|^2 = 1$ and $w_2 = 0$, then $\mu_x = 0$, $\mu_y = 0$, $\mu_z = e\hbar/(2mc)$; and, if $w_1 = 0$ and $|w_2|^2 = 1$, then $\mu_x = 0$, $\mu_y = 0$, $\mu_z = -e\hbar/(2mc)$. These results are analogous to (3.19).

## 4. The anomalous electron magnetic moment

We are coming now to the next approximation to $\psi$ upon writing $\psi = \psi_0 + \psi_1$ where the zeroth approximation $\psi_0$ is given by (3.21). The first correction to the potential energy of (3.20) will be

$$\Pi_1 = \alpha \int_{(\infty)} \left(\bar{\psi}_0 \gamma^\mu \psi_1 + \bar{\psi}_1 \gamma^\mu \psi_0\right) A_\mu^{ext} dV. \tag{4.1}$$

The bispinor $\psi_1$ should, however, be renormalized, which is seen, for example, from § 33 of [3]. We carry out the renormalization in parallel with Schweber [5, Sec. 15c] adding the factor ½:

$$\Pi_1 = \tfrac{1}{2}\alpha \int_{(\infty)} \left( \overline{\psi}_0 \gamma^\mu \psi_1 + \overline{\psi}_1 \gamma^\mu \psi_0 \right) A_\mu^{\text{ext}} dV . \tag{4.2}$$

The equation for $\psi_1$ follows from (2.5) with $A_\mu^{\text{ext}} = 0$ since $A_\mu^{\text{ext}}$ is taken into account in (4.2):

$$i\gamma^\mu \frac{\partial \psi_1}{\partial x^\mu} - \psi_1 = \alpha A_\mu \gamma^\mu \psi_0 . \tag{4.3}$$

The bispinor $\psi_1$ is of the form

$$\psi_1 = \psi_1(\mathbf{r}) e^{-i\varepsilon t} . \tag{4.4}$$

In the following we shall sometimes omit the argument $\mathbf{r}$ of $\psi_1(\mathbf{r})$ for brevity sake. Substituting (4.4) into (4.3) yields

$$i\alpha^k \frac{\partial \psi_1}{\partial x^k} + (\varepsilon - \beta)\psi_1 = \alpha A_\mu \gamma^0 \gamma^\mu \psi_0(\mathbf{r}) . \tag{4.5}$$

where $\beta = \gamma^0$ as is customary.

The solution to this equation can be written as

$$\psi_1(\mathbf{r}) = \alpha \int_{(\infty)} G(\mathbf{r}-\mathbf{r}') A_\mu(\mathbf{r}') \gamma^0 \gamma^\mu \psi_0(\mathbf{r}') dV' , \tag{4.6}$$

where $G(\mathbf{r}-\mathbf{r}')$ is a Green's function satisfying the equation

$$\left( i\alpha^k \frac{\partial}{\partial x^k} + \varepsilon - \beta \right) G(\mathbf{r}-\mathbf{r}') = \delta(\mathbf{r}-\mathbf{r}') . \tag{4.7}$$

It is well known that the Green's function satisfying the equation $\nabla^2 G_0 + p^2 G_0 = \delta(\mathbf{r}-\mathbf{r}')$ is

$$G_0(\mathbf{r}-\mathbf{r}') = -\frac{1}{4\pi} \frac{e^{\pm ip|\mathbf{r}-\mathbf{r}'|}}{|\mathbf{r}-\mathbf{r}'|} . \tag{4.8}$$

It is not difficult to ascertain now that the Green's function of Eq. (4.7) will be

$$G(\mathbf{r}-\mathbf{r}') = \frac{1}{4\pi} \left( i\alpha^k \frac{\partial}{\partial x^k} - \varepsilon - \beta \right) \frac{e^{ip|\mathbf{r}-\mathbf{r}'|}}{|\mathbf{r}-\mathbf{r}'|} , \tag{4.9}$$

where $\mathbf{p}^2 = \varepsilon^2 - 1$, $p = |\mathbf{p}|$ and we take the upper sign in (4.8). We need the Fourier transform of $G(\mathbf{r}-\mathbf{r}')$ for what follows. With use made of Eq. (3.26) we obtain

$$G(\mathbf{k}) = \frac{\alpha^l k_l + \varepsilon + \beta}{\mathbf{p}^2 - \mathbf{k}^2 + i\xi} , \tag{4.10}$$

where $\xi \to 0$. The quantity $\xi$ is added inasmuch as the denominator would vanish when $\mathbf{p}^2 = \mathbf{k}^2$. This is a usual practice in SQED.

Equation (4.2) contains the Dirac conjugate bispinor $\overline{\psi}_1 = \psi_1^* \gamma^0$. This bispinor can be found starting from (4.6) with the result

10$$\overline{\psi}_1(\mathbf{r}) = \alpha \int_{(\infty)} A_\mu(\mathbf{r}')\overline{\psi}_0(\mathbf{r}')\gamma^\mu G^+(\mathbf{r}-\mathbf{r}')\gamma^0 dV', \qquad (4.11)$$

where the superscript + denotes a Hermitian conjugate quantity. As long as the matrices $\alpha^l$ and $\beta$ are Hermitian, we have from (4.10) that

$$G^+(\mathbf{k}) = \frac{\alpha^l k_l + \varepsilon + \beta}{\mathbf{p}^2 - \mathbf{k}^2 - i\xi}. \qquad (4.12)$$

We place $\psi_1(\mathbf{r})$ of (4.6) and $\overline{\psi}_1(\mathbf{r})$ of (4.11) in Eq. (4.2) together with the spatial part of $\psi$ (3.21) for $\psi_0$ with slightly different $\varepsilon'$, $\mathbf{p}'$ and $\varepsilon$, $\mathbf{p}$ and we put $\varepsilon' = \varepsilon$ at once as in (3.28). Integration over $V$ and $V'$ leads to the Fourier transforms of $G(\mathbf{r} - \mathbf{r}')$ and $A_\nu(\mathbf{r})$ with the result

$$\Pi_1 = \frac{\alpha^2}{4\varepsilon(2\pi)^3}\int_{(\infty)} A_\mu^{\text{ext}}(\mathbf{k})\overline{u}(\mathbf{p}')\left[\gamma^\mu G(\mathbf{p}'-\mathbf{k})A_\nu(-\mathbf{k})\gamma^0\gamma^\nu + A_\nu^*(\mathbf{k})\gamma^\nu G^+(\mathbf{p}+\mathbf{k})\gamma^0\gamma^\mu\right]u(\mathbf{p})d\mathbf{k}. \quad (4.13)$$

The electromagnetic field described by $A_\mu(\mathbf{r})$ in (4.6) or $A_\nu(\mathbf{k})$ in (4.13) is the electromagnetic field due to the electron. The electron creates an electric field and a magnetic field owing to its spin. The magnetic field can, however, be neglected in the present calculations because it gives corrections of the next order in $\alpha$. The potential of the electric field in the same approximation is $\varphi = 1/r$. The Fourier transform of the Coulomb potential $1/r$ is well-known:

$$A_0(\mathbf{k}) = \frac{4\pi}{\mathbf{k}^2 + \xi^2} \qquad (4.14)$$

with $\xi \to 0$.

It is much more convenient to work with the external scalar potential $\varphi^{\text{ext}} = A_0^{\text{ext}}$ rather than with the vector potential $\mathbf{A}^{\text{ext}}$. For example, Sokolov [6, § 44] starts with the scalar potential and only afterwards he passes on to the vector potential. Akhiezer and Berestetskii [7, § 44] work with the electric and magnetic external fields simultaneously and only at the end they preserve the magnetic field alone. If we retain the summands with $\mu = \nu = 0$ alone in the sums of Eq. (4.13), the equation will be markedly simplified. Besides, to further simplify formulae we put $\mathbf{p}' = \mathbf{p}$ at once in the square brackets of (4.13) because we intend to put $\mathbf{p}' = \mathbf{p}$ at the end of calculations. Upon substituting (4.14) we shall obtain

$$\Pi_1 = \frac{\alpha^2}{8\varepsilon\pi^2}\int_{(\infty)}\frac{\varphi^{\text{ext}}(\mathbf{k})}{\mathbf{k}^2 + \xi^2}u^*(\mathbf{p}')\left[G(\mathbf{p}-\mathbf{k}) + G^+(\mathbf{p}+\mathbf{k})\right]u(\mathbf{p})d\mathbf{k}. \qquad (4.15)$$

We imply that the external field varies only slowly in space, in which case $\mathbf{k}$ is small (in the following we shall take the limit $\mathbf{k} \to 0$ in parallel with SQED). In this situation the sum $G(\mathbf{p} - \mathbf{k}) + G^+(\mathbf{p} + \mathbf{k})$ that figures in (4.15) and computed with the help of (4.10) and (4.12) acquires the following simple form

$$G(\mathbf{p}-\mathbf{k}) + G^+(\mathbf{p}+\mathbf{k}) = \frac{2\alpha\mathbf{k}}{2\mathbf{p}\mathbf{k}+i\xi}, \tag{4.16}$$

if $k^2$ is deleted. One should take into account the relation $u^*(\mathbf{p})\alpha u(\mathbf{p}) = \mathbf{p}u^*(\mathbf{p})u(\mathbf{p})/\varepsilon$ which follows from (3.22). The integral in (4.15) can now be readily calculated with use made of spherical coordinates upon assuming a spherical symmetry of $\varphi^{ext}(\mathbf{k})$ in $\mathbf{k}$-space. We can now put $\xi = 0$ with the result

$$\Pi_1 = \frac{\alpha^2}{2\varepsilon^2\pi}\int_0^\infty u^*(\mathbf{p}')\varphi^{ext}(k)u(\mathbf{p})dk = \frac{\alpha^2}{4\varepsilon^2\pi}\int_{-\infty}^\infty u^*(\mathbf{p}')\varphi^{ext}(k)u(\mathbf{p})dk. \tag{4.17}$$

The scalar potential $\varphi^{ext}$ is contained in the relativistic invariant $\gamma^\mu A_\mu^{ext} = \gamma^0\varphi^{ext} - \gamma\mathbf{A}^{ext} = \gamma^0(\varphi^{ext} - \alpha\mathbf{A}^{ext})$ which enters in the starting potential energy of (4.2). Therefore, $\varphi^{ext}$ in (4.17) can be replaced by $-\alpha\mathbf{A}^{ext}$ in parallel with Sokolov [6]:

$$\Pi_1 = -\frac{\alpha^2}{4\varepsilon^2\pi}\int_{-\infty}^\infty u^*(\mathbf{p}')\alpha\mathbf{A}^{ext}(k)u(\mathbf{p})dk. \tag{4.18}$$

As mentioned above we imply that the external field varies only slowly in space, in which case $\mathbf{A}^{ext}$ is factually proportional to a delta function, namely, $\mathbf{A}^{ext}(k) = A_0^{ext}(k)\delta(k)$. The integration in (4.18) can now be carried out at once. Omitting the subscript 0 one has

$$\Pi_1 = -\frac{\alpha^2}{4\varepsilon^2\pi}u^*(\mathbf{p}')\alpha\mathbf{A}^{ext}(k)u(\mathbf{p}), \tag{4.19}$$

implying that $k \to 0$. Upon comparing this with (3.29) and recalling that $\varepsilon = 1$ in the non-relativistic limit we see that $\Pi_1$ contains an extra factor equal to $\alpha/(2\pi)$. Therefore the resulting electron magnetic moment in this approximation will be

$$\mu = \frac{e\hbar}{2mc}\left(1 + \frac{\alpha}{2\pi}\right). \tag{4.20}$$

Thus we have obtained the Schwinger correction $\alpha/(2\pi)$ in NAQED.

### 5. The Lamb shift

SQED contains an important procedure called the renormalization of electron mass. As mentioned above the electron mass is absent in the starting equations of NAQED; it appears in NAQED in the non-relativistic limit alone. At the same time, the ordinary relation between the electron mass $m$ and the electron Compton wavelength $\lambdabar$, $m = \hbar/\lambdabar c$, does not hold in NAQED. The observed mass $m$ of a free electron and $\lambdabar$ are connected by Eq. (5.2) of I

$$m = \frac{\hbar}{c\lambdabar}\left\{1 - \alpha c_0 + \alpha \int\limits_{(\infty)}\left[\frac{1}{8\pi}\left(\tilde{E}^2 + \tilde{H}^2\right) - \overline{\varphi}\left(\tilde{\psi}_1^*\tilde{\psi}_1 - \tilde{\psi}_2^*\tilde{\psi}_2\right)\right]d\tilde{V}\right\}. \tag{5.1}$$

We must put $c_0 = 0$ here because the scalar potential $\varphi$ of I includes an energy and does not vanish at infinity giving rise to $c_0$ while the present $\varphi$ vanishes at infinity. Seeing that $\mathbf{E} = -\nabla\varphi$ in a stationary state and using Green's theorem and Maxwell's equations one has

$$\int\limits_{(\infty)}\frac{\tilde{E}^2}{8\pi}dV = \tfrac{1}{2}\int\limits_{\infty}\overline{\varphi}\left(\tilde{\psi}_1^*\tilde{\psi}_1 - \tilde{\psi}_2^*\tilde{\psi}_2\right)d\tilde{V}. \tag{5.2}$$

Upon putting $\psi_1 = \psi$, $\psi_2 = 0$ and removing the tildes as above, setting $c_0 = 0$ and using (5.2) we recast Eq. (5.1) conveniently as

$$m = \frac{\hbar}{c\lambdabar}\left\{1 + \alpha\int\limits_{(\infty)}\left[\zeta_H\frac{H^2}{8\pi} + \zeta_\varphi\varphi\psi^*\psi\right]dV\right\}^{-1}. \tag{5.3}$$

Here we have inserted unknown constant coefficients $\zeta_H$ and $\zeta_\varphi$ because in this section we imply that the electron is not free but it resides in the Coulomb field of an atom. The coefficients will be found later.

Instead of $\lambdabar$ we introduce an effective $\lambdabar_R$ in order to have the habitual relation

$$m = \frac{\hbar}{c\lambdabar_R}. \tag{5.4}$$

We define now dimensionless quantities according to (2.4) but we write $\lambdabar_R$ instead of $\lambdabar$. In this case the energy $E$ will be

$$E = \frac{c\hbar}{\lambdabar_R}\tilde{E} = mc^2\tilde{E}, \tag{5.5}$$

in which (5.4) is taken into account. Thus we shall have the standard relation between the energy $E$ and the rest mass $m$.

This procedure can be called the $\lambdabar$-renormalization by analogy with the mass renormalization in SQED. As opposed to SQED, however, the $\lambdabar$-renormalization does not involve divergent expressions. All quantities that figure in the $\lambdabar$-renormalization are finite and strictly defined.

Comparing (5.3) and (5.4) yields

$$\frac{\lambdabar_R}{\lambdabar} = 1 + \alpha\delta\lambda, \qquad \delta\lambda = \int\limits_{(\infty)}\left[\zeta_H\frac{H^2}{8\pi} + \zeta_\varphi\varphi\psi^*\psi\right]dV. \tag{5.6}$$

It should be emphasized that the quantity $\lambdabar$ enters in the starting equation of (2.1), and not $\lambdabar_R$. When we use the dimensionless quantities of (2.4) with $\lambdabar_R$, an extra term appears in (2.5) owing to (5.6):

$$i\gamma^\mu \frac{\partial \psi}{\partial x^\mu} - \alpha(A_\mu + A_\mu^{\text{ext}})\gamma^\mu \psi - (1+\alpha\delta\lambda)\psi = 0. \tag{5.7}$$

It can be seen from this equation that it is just Eq. (5.3) which is appropriate for the $\lambdabar$-renormalization in the present situation. The extra term with $\delta\lambda$ is analogous to $\delta m$ in SQED. At the same time, $\delta m$ of SQED diverges and requires a "naked" mass $m_0$ for compensation whereas $\delta\lambda$ is finite in view of (5.6).

We have for a stationary state with energy $\varepsilon$ that $\psi = \psi(\mathbf{r})e^{-i\varepsilon t}$ [we shall omit the argument $\mathbf{r}$ in $\psi(\mathbf{r})$], so that Eq. (5.7) becomes

$$i\alpha^k \frac{\partial \psi}{\partial x^k} - \alpha(A_\mu + A_\mu^{\text{ext}})\gamma^0\gamma^\mu \psi + [\varepsilon - (1+\alpha\delta\lambda)\beta]\psi = 0. \tag{5.8}$$

We regard the terms with $A_\mu$ and $\delta\lambda$ in (5.8) as perturbations. The starting equation without the perturbations is

$$i\alpha^k \frac{\partial \psi^{(0)}}{\partial x^k} - \alpha A_\mu^{\text{ext}}\gamma^0\gamma^\mu \psi^{(0)} + (\varepsilon - \beta)\psi^{(0)} = 0. \tag{5.9}$$

We imply that the electron is in the Coulomb field of an atom or ion with charge $-Ze$. In this case the dimensionless potential will be $A_0^{\text{ext}} \equiv \varphi^{\text{ext}} = -Z/r$ and Eq. (5.9) will take the form

$$i\alpha^k \frac{\partial \psi^{(0)}}{\partial x^k} + \frac{Z\alpha}{r}\psi^{(0)} + (\varepsilon - \beta)\psi^{(0)} = 0. \tag{5.10}$$

We borrow solutions to this equation from Ref. [3, § 36]. As mentioned above it should be kept in mind that, when applying formulae of [3] for our dimensionless quantities, one should put $m = 1$ for the electron mass. The Lamb shift concerns the electron states $2s_{1/2}$ and $2p_{1/2}$. The $2s_{1/2}$ state wherein $n = 2$, $l = 0$, $j = m = \frac{1}{2}$, $n_r = 1$, $\kappa = -1$, is described by the bispinor

$$\psi_1^{(0)} = \begin{pmatrix} \eta_1 \\ \chi_1 \end{pmatrix}, \qquad \eta_1 = \frac{1}{\sqrt{4\pi}}\begin{pmatrix} f_1(r) \\ 0 \end{pmatrix}, \qquad \chi_1 = \frac{-i}{\sqrt{4\pi}}\begin{pmatrix} g_1(r)\cos\vartheta \\ g_1(r)\sin\vartheta\, e^{i\varphi} \end{pmatrix}, \tag{5.11}$$

where the functions $f_1(r)$ and $g_1(r)$ are

$$f_1(r) = \sqrt{1+\varepsilon}\, A_1 r^{\gamma-1} e^{-\lambda r}\left[\frac{Z\alpha}{\lambda} - \left(\frac{Z\alpha}{\lambda}+1\right)\frac{2\lambda r}{2\gamma+1}\right],$$

$$g_1(r) = -\sqrt{1-\varepsilon}\, A_1 r^{\gamma-1} e^{-\lambda r}\left[\frac{Z\alpha}{\lambda} + 2 - \left(\frac{Z\alpha}{\lambda}+1\right)\frac{2\lambda r}{2\gamma+1}\right] \tag{5.12}$$

with

$$A_1^2 = \frac{(2\gamma+1)(2\lambda)^{2\gamma+2}}{8Z\alpha\left(\frac{Z\alpha}{\lambda}+1\right)\Gamma(2\gamma+1)}. \tag{5.13}$$

Hereinafter

$$\gamma = \sqrt{\kappa^2 - (Z\alpha)^2}, \quad \lambda = \sqrt{1-\varepsilon^2}. \tag{5.14}$$

The $2p_{1/2}$ state for which $n = 2$, $l = 1$, $j = m = \frac{1}{2}$, $n_r = 1$, $\kappa = 1$, is described by the bispinor

$$\psi_2^{(0)} = \begin{pmatrix} \eta_2 \\ \chi_2 \end{pmatrix}, \quad \eta_2 = \frac{-i}{\sqrt{4\pi}}\begin{pmatrix} f_2(r)\cos\vartheta \\ f_2(r)\sin\vartheta e^{i\varphi} \end{pmatrix}, \quad \chi_2 = \frac{-1}{\sqrt{4\pi}}\begin{pmatrix} g_2(r) \\ 0 \end{pmatrix}, \tag{5.15}$$

where the functions $f_2(r)$ and $g_2(r)$ are

$$f_2(r) = \sqrt{1+\varepsilon}\,A_2 r^{\gamma-1} e^{-\lambda r}\left[\frac{Z\alpha}{\lambda} - 2 - \left(\frac{Z\alpha}{\lambda}-1\right)\frac{2\lambda r}{2\gamma+1}\right],$$

$$g_2(r) = -\sqrt{1-\varepsilon}\,A_2 r^{\gamma-1} e^{-\lambda r}\left[\frac{Z\alpha}{\lambda} - \left(\frac{Z\alpha}{\lambda}-1\right)\frac{2\lambda r}{2\gamma+1}\right] \tag{5.16}$$

with

$$A_2^2 = \frac{(2\gamma+1)(2\lambda)^{2\gamma+2}}{8Z\alpha\left(\frac{Z\alpha}{\lambda}-1\right)\Gamma(2\gamma+1)}. \tag{5.17}$$

The above functions satisfy the normalization condition

$$\int_0^\infty \left(f_i^2 + g_i^2\right) r^2 dr = 1, \quad i = 1, 2. \tag{5.18}$$

The energy of these degenerate levels minus the rest energy is

$$\varepsilon_0 - 1 = -\frac{(Z\alpha)^2}{8} - \frac{5(Z\alpha)^4}{128} - \frac{21(Z\alpha)^6}{1024} - \ldots \tag{5.19}$$

The Lamb shift in a first approximation is treated in SQED in the non-relativistic limit. By the way, the mass in NAQED appears in the non-relativistic limit as well, so that Eqs. (5.3) and (5.4) are meaningful in NAQED in this limit alone. If we represent the bispinor $\psi$ as $\psi = \begin{pmatrix}\eta\\\chi\end{pmatrix}$, Eq. (5.9) in the non-relativistic limit transforms into [3, § 33]

$$\tfrac{1}{2}\nabla^2\eta + \left(\bar{\varepsilon} + \frac{Z\alpha}{r}\right)\eta - \alpha(\varphi\eta - \mathbf{A}\boldsymbol{\sigma}\chi + \delta\lambda\eta) = 0, \tag{5.20}$$

where $\bar{\varepsilon} = \varepsilon - 1$ is the non-relativistic energy. We have written down only the equation for $\eta$ because the equation for $\chi$ is not needed for our study. The perturbation operator in Eq. (5.20) together with the functions on which it acts is

$$V = -\alpha(\varphi\eta - \mathbf{A}\boldsymbol{\sigma}\chi + \delta\lambda\eta). \tag{5.21}$$

Seeing that the states $2s_{1/2}$ and $2p_{1/2}$ under consideration are degenerate from the viewpoint of the starting equation of (5.10) we turn to Landau and Lifshitz [8, § 39] in order to apply perturbation theory in the case where there are degenerate eigenvalues. The equations for the correction $\varepsilon_1$ to the energy in this case are

$$(\varepsilon_1 - V_{11})c_1^{(0)} = V_{12}c_2^{(0)}, \tag{5.22}$$

$$V_{21}c_1^{(0)} = (\varepsilon_1 - V_{22})c_2^{(0)}, \tag{5.23}$$

where

$$V_{kn} = -\alpha \int_{(\infty)} \eta_k^* (\varphi\eta_n - \mathbf{A}\boldsymbol{\sigma}\chi_n + \delta\lambda\eta_n) \, dV \tag{5.24}$$

on account of (5.21).

Equation (5.20) is in fact the Schrödinger equation with a perturbation. At the same time it should be emphasized that when calculating $V_{kn}$ of (5.24) one should use the exact wave functions of (5.11) and (5.15) satisfying the Dirac equation since the perturbation operator of (5.21) contains a magnetic field created by the electron owing to its spin whereas this field can be found with the help of the Dirac equation alone.

We begin with

$$V_{12} = V_{12}^{(1)} + V_{12}^{(2)}, \quad V_{12}^{(1)} = -\alpha \int_{(\infty)} \eta_1^*(\varphi + \delta\lambda)\eta_2 \, dV, \quad V_{12}^{(2)} = \alpha \int_{(\infty)} \eta_1^* \mathbf{A}\boldsymbol{\sigma}\chi_2 \, dV. \tag{5.25}$$

The scalar potential $\varphi$ created by the electron in an atom is spherically symmetric. When we substitute Eqs. (5.11) and (5.15) into $V_{12}^{(1)}$, we shall see that $V_{12}^{(1)} = 0$ thanks to integration over the angle $\theta$. This is obvious in advance inasmuch as the eigenfunctions $\psi_1^{(0)}$ and $\psi_2^{(0)}$ are mutually orthogonal. The vector potential $\mathbf{A}$ due to the electron in the atom under consideration has only one component $A_\varphi$ [9]. Therefore

$$\mathbf{A}\boldsymbol{\sigma}\begin{pmatrix}\chi^{(1)}\\\chi^{(2)}\end{pmatrix} = iA_\varphi\begin{pmatrix}-\chi^{(2)}e^{-i\varphi}\\\chi^{(1)}e^{i\varphi}\end{pmatrix}. \tag{5.26}$$

It is not now difficult to verify that $V_{12}^{(2)} = 0$ too. As a result, $V_{12} = 0$. Analogously, $V_{21} = 0$ as well.

As long as $V_{12} = V_{21} = 0$ the set of equations (5.22) and (5.23) admits two solutions

$$1) \quad c_1^{(0)} = 1, \quad c_2^{(0)} = 0, \quad \varepsilon_1 = V_{11}, \tag{5.27}$$

$$2) \quad c_1^{(0)} = 0, \quad c_2^{(0)} = 1, \quad \varepsilon_2 = V_{22}. \tag{5.28}$$

We start with $V_{11}$ which gives a correction $\varepsilon_1$ to the energy of (5.19) in the $2s_{1/2}$ state by (5.27). We have from (5.24) that

$$V_{11} = V_{11}^{(1)} + V_{11}^{(2)} + V_{11}^{(3)}, \quad V_{11}^{(1)} = -\alpha \int_{(\infty)} \eta_1^* \varphi \eta_1 \, dV, \quad V_{11}^{(2)} = -\alpha \delta\lambda \int_{(\infty)} \eta_1^* \eta_1 \, dV, \quad V_{11}^{(3)} = \alpha \int_{(\infty)} \eta_1^* \mathbf{A}\boldsymbol{\sigma}\chi_1 \, dV.$$

(5.29)

As long as we carry out our calculations in the non-relativistic approximation all constant quantities should be expanded in power series in $\alpha$. It follows from (5.14) and (5.19) that the power series will contain only even powers of $Z\alpha$. At the same time, quantities of the type $(Z\alpha r)^n$ must remain intact inasmuch as, after integration over $r$ with $e^{-Z\alpha r}$, they will yields numbers independent of $\alpha$. We obtain from (5.12)–(5.14) in our approximation that

$$f_1(r) = \left(\frac{Z\alpha}{2}\right)^{3/2}(2 - Z\alpha r)e^{-\frac{1}{2}Z\alpha r}, \quad g_1(r) = -\frac{1}{2}\left(\frac{Z\alpha}{2}\right)^{5/2}(4 - Z\alpha r)e^{-\frac{1}{2}Z\alpha r}.$$

(5.30)

The normalization conditions follow from (5.18):

$$\int_{(\infty)} \eta_1^* \eta_1 dV = \int_0^\infty f_1^2 r^2 dr = 1 - \frac{(Z\alpha)^2}{16}, \quad \int_0^\infty g_1^2 r^2 dr = \frac{(Z\alpha)^2}{16},$$

(5.31)

although one would have directly from (5.30) that $\int_0^\infty f_1^2 r^2 dr = 1$.

The scalar potential $\varphi$ that figures in $V_{11}^{(1)}$ of (5.29) is created by the electron in our atom. The potential is considered in Appendix A and is given by Eq. (A.5). Calculation with the potential and $f_1(r)$ of (5.30) yields

$$V_{11}^{(1)} = -\frac{77}{512} Z\alpha^2 + c_1 \alpha (Z\alpha)^3.$$

(5.32)

The constant $c_1$ can be found as a result of tedious but straightforward calculations with use made of the exact functions of (5.12) but the concrete value of $c_1$ is not required for what follows.

We recast $\delta\lambda$ of (5.6) as

$$\delta\lambda = \zeta_H \delta\lambda_H + \zeta_\varphi \delta\lambda_\varphi, \quad \delta\lambda_H = \int_{(\infty)} \frac{H^2}{8\pi} dV, \quad \delta\lambda_\varphi = \int_{(\infty)} \varphi \psi^* \psi dV.$$

(5.33)

In a first approximation $\delta\lambda_\varphi = -V_{11}^{(1)}/\alpha$, so that by analogy with (5.32)

$$\delta\lambda_\varphi = \frac{77}{512} Z\alpha + c_2 (Z\alpha)^3.$$

(5.34)

The concrete value of $c_2$ will not be needed for our calculations.

Seeing that $\mathbf{H} = \text{curl}\mathbf{A}$, to calculate $\delta\lambda_H$ we must know the vector potential $\mathbf{A}$ created by the electron in our atom. The vector potential in this case is studied in detail in Ref. [9]. The vector

**A** has only one component $A_\varphi$. In the state $2s_{1/2}$, $A_\varphi = \zeta(r)\sin\theta$ with $\zeta(r)$ given by Eq. (3.23) of [9]. In our non-relativistic approximation that equation reduces to

$$\zeta(r) = \frac{1}{2r^2} - \frac{1}{2r^2}\left(1 + Z\alpha r + \frac{1}{2}(Z\alpha r)^2 + \frac{1}{8}(Z\alpha r)^4\right)e^{-Z\alpha r}. \tag{5.35}$$

The intensity of the magnetic field **H** in this case has two nonzero components

$$H_r = \frac{2\zeta}{r}\cos\theta, \quad H_\theta = -\left(\frac{\zeta}{r} + \frac{d\zeta}{dr}\right)\sin\theta; \quad H^2 = H_r^2 + H_\theta^2. \tag{5.36}$$

Substituting these into $\delta\lambda_H$ of (5.33) yields

$$\delta\lambda_H = a_H(Z\alpha)^3, \quad a_H = \frac{1}{36}\left(\ln 2 - \frac{685}{1024}\right). \tag{5.37}$$

In view of (5.31) and (5.34) the main terms in the second summand in $V_{11}$ of (5.29) will now be

$$V_{11}^{(2)} = -\alpha\left[\frac{77}{512}\zeta_\varphi Z\alpha + (\zeta_H a_H + \zeta_\varphi c_3)(Z\alpha)^3\right], \tag{5.38}$$

where $c_3$ is a new constant which contains the above constant $c_2$ and an extra constant due to the second term in $\int_{(\infty)} \eta_1^*\eta_1 dV$ of (5.31).

It remains for us to compute the last summand in $V_{11}$ of (5.29). With use made of (5.11) and (5.26) we obtain

$$V_{11}^{(3)} = \frac{\alpha}{4\pi}\int_{(\infty)} f_1(r)g_1(r)A_\varphi(r,\theta)\sin\theta dV = \frac{2\alpha}{3}\int_0^\infty f_1(r)g_1(r)\zeta(r)r^2 dr, \tag{5.39}$$

where the relationship $A_\varphi = \zeta(r)\sin\theta$ was used. Substituting Eqs. (5.30) and (5.35) results in

$$V_{11}^{(3)} = -\frac{5\alpha(Z\alpha)^3}{6144}. \tag{5.40}$$

Upon combining Eqs. (5.32), (5.38) and (5.40) we are in position to calculate $V_{11}$ of (5.29) and thereby to find the correction $\varepsilon_1$ to the energy of the $2s_{1/2}$ level by (5.27):

$$\varepsilon_1 = -\alpha(1+\zeta_\varphi)\frac{77}{512}Z\alpha + \alpha\left(c_1 - \zeta_H a_H - \zeta_\varphi c_3 - \frac{5}{6144}\right)(Z\alpha)^3 + O(\alpha^6). \tag{5.41}$$

The first term in $\varepsilon_1$ is of the same order $\alpha^2$ as the first term at the right of Eq. (5.19), the term being the non-relativistic energy of the electron. This energy cannot be changed by a small perturbation, so that $\zeta_\varphi = -1$ in the $\lambda$-renormalization of (5.6). Analogously we put

$$\zeta_H = \frac{1}{a_H}\left(c_1 - \zeta_\varphi c_3 - \frac{5}{6144}\right), \tag{5.42}$$

and the second term in $\varepsilon_1$ disappears as well. Because of this it is necessary to consider the next approximation in perturbation theory.

The second correction $\varepsilon_1^{(2)}$ to the energy of degenerate levels is given by Landau and Lifshitz [8] in Problem 2 of § 39:

$$\varepsilon_1^{(2)} = {\sum_m}' \frac{V_{1m} V_{m1}}{\varepsilon_1^{(0)} - \varepsilon_m^{(0)}}, \tag{5.43}$$

where the prime denotes that $m \neq 1, 2$. It should be remarked that the eigenfunctions remain as before by (5.27). According to (5.24)

$$V_{1m} = -\alpha \int_{(\infty)} \eta_1^* (\varphi \eta_m - \mathbf{A}\boldsymbol{\sigma}\chi_m + \delta\lambda \eta_m) dV, \tag{5.44}$$

and an analogous expression for $V_{m1}$. The last term in (5.44) vanishes because $\eta_1$ and $\eta_m$ are mutually orthogonal if $m \neq 1, 2$. The first term, by analogy with $V_{11}^{(1)}$ of (5.29) and (5.32), is proportional to $Z\alpha^2$; the second term, by analogy with $V_{11}^{(3)}$ of (5.29) and (5.40), is proportional to $\alpha(Z\alpha)^3$. The difference in the denominator of (5.43) is proportional to $(Z\alpha)^2$ since all energy levels are proportional to $(Z\alpha)^2$ in our atom as in (5.19). As a result, the correction $\varepsilon_1^{(2)}$ of (5.43) will contain terms of the orders $\alpha^2$, $\alpha^4$, $\alpha^6$, etc. because the power series in our case include even powers of $\alpha$ alone as mentioned above. When $\varepsilon_1^{(2)}$ is added to $\varepsilon_1$ of (5.41), this will result in a renormalization of the constants $\zeta_\varphi$ and $\zeta_H$ which can again be so chosen that the first two terms in (5.41) vanish as before.

The constant $\zeta_H$ may, however, contain a small addition proportional to $Z\alpha$, and we replace $\zeta_H$ by $\zeta_H + \zeta_1 Z\alpha$, in which case Eq. (5.41) gives

$$\varepsilon_1 = -\alpha a_H \zeta_1 (Z\alpha)^4. \tag{5.45}$$

Thus we have obtained the level shift $\varepsilon_1$ which is of the same order $\alpha(Z\alpha)^4$ as the Lamb shift including the number $Z$. At the same time the constant $\zeta_1$ cannot be found with the help of the method employed in the present section. That is the reason why one in SQED does not resort to perturbation theory for degenerated levels and uses other methods for calculating the Lamb shift.

The point is that excited states of an atom, in particular the states $2s_{1/2}$ and $2p_{1/2}$ under consideration, are not strictly stationary, but only quasi-stationary, giving rise to finite level widths. A complex energy value can be assigned to such states where the imaginary part is proportional to the decay probability of the state. The decay of the state is a nonstationary process whereas our consideration in this paper is restricted to stationary states and we implied that all energy values are real. The methods developed in SQED enable one to consider the decay of the quasi-stationary states as well. For example, Akhiezer and Berestetskii [7, § 53] calculate level shifts and level widths simultaneously. If we take the results of [7] concerning the Lamb shift, we shall get that $\zeta_1 \approx -595$.

We are coming now to the state $2p_{1/2}$. We obtain from (5.16) and (5.17) in our non-relativistic approximation that

$$f_2(r) = -(Z\alpha)^{5/2} \frac{r}{2\sqrt{6}} e^{-\frac{1}{2}Z\alpha r}, \quad g_2(r) = -(Z\alpha)^{5/2} \frac{\sqrt{3}}{8\sqrt{2}} \left(2 - \frac{Z\alpha r}{3}\right) e^{-\frac{1}{2}Z\alpha r}. \tag{5.46}$$

The normalization conditions are

$$\int_{(\infty)} \eta_2^* \eta_2 dV = \int_0^\infty f_2^2 r^2 dr = 1 - \frac{(Z\alpha)^2}{16}, \quad \int_0^\infty g_2^2 r^2 dr = \frac{(Z\alpha)^2}{16}, \tag{5.47}$$

although one would have directly from (5.46) that $\int_0^\infty f_2^2 r^2 dr = 1$.

According to (5.28) to obtain a correction $\varepsilon_2$ to the energy of (5.19) in the $2p_{1/2}$ state one must calculate the quantity $V_{22}$ whose formula follows from (5.24), namely,

$$V_{22} = V_{22}^{(1)} + V_{22}^{(2)} + V_{22}^{(3)}, \, V_{22}^{(1)} = -\alpha \int_{(\infty)} \eta_2^* \varphi \eta_2 \, dV, \, V_{22}^{(2)} = -\alpha\delta\lambda \int_{(\infty)} \eta_2^* \eta_2 \, dV, \, V_{22}^{(3)} = \alpha \int_{(\infty)} \eta_2^* \mathbf{A}\boldsymbol{\sigma}\chi_2 \, dV$$
$$. \tag{5.48}$$

Calculation with the potential of (A.6) and $f_2(r)$ of (5.46) yields

$$V_{22}^{(1)} = -\frac{93}{512} Z\alpha^2 + \bar{c}_1 \alpha(Z\alpha)^3. \tag{5.49}$$

The concrete value of $\bar{c}_1$ and of other similar constants $\bar{c}_i$ will not be needed for what follows.

We take $\delta\lambda$ as written in (5.33). In a first approximation $\delta\lambda_\varphi = -V_{22}^{(1)}/\alpha$, so that by analogy with (5.34) and (5.49) we obtain

$$\delta\lambda_\varphi = \frac{93}{512} Z\alpha + \bar{c}_2 (Z\alpha)^3. \tag{5.50}$$

To calculate $\delta\lambda_H$ it is necessary to know the vector potential $\mathbf{A}$. In the $2p_{1/2}$ state as in the $2s_{1/2}$ state the vector $\mathbf{A}$ has only one component $A_\varphi = \zeta(r)\sin\theta$ with $\zeta(r)$ given by Eq. (3.31) of [9]. In our non-relativistic approximation that equation reduces to

$$\zeta(r) = \frac{1}{6r^2} - \frac{1}{6r^2}\left(1 + Z\alpha r + \frac{1}{2}(Z\alpha r)^2 - \frac{1}{8}(Z\alpha r)^4\right) e^{-Z\alpha r}. \tag{5.51}$$

Calculating $\delta\lambda_H$ with use made of (5.36) gives

$$\delta\lambda_H = \bar{a}_H (Z\alpha)^3, \quad \bar{a}_H = \frac{1}{162}\left(\ln 2 + \frac{737}{256}\right). \tag{5.52}$$

The main terms in the second summand in $V_{22}$ of (5.48) will now be

$$V_{22}^{(2)} = -\alpha\left[\frac{93}{512}\zeta_\varphi Z\alpha + \left(\zeta_H \bar{a}_H + \zeta_\varphi \bar{c}_3\right)(Z\alpha)^3\right], \tag{5.53}$$

where $\bar{c}_3$ is a new constant.

With use made of (5.15) and (5.26) the last summand in $V_{22}$ of (5.48) rearranges to

$$V_{22}^{(3)} = \frac{\alpha}{4\pi} \int_{(\infty)} f_2(r)g_2(r)A_\varphi(r,\theta)\sin\theta dV = \frac{2\alpha}{3}\int_0^\infty f_2(r)g_2(r)\zeta(r)r^2 dr, \quad (5.54)$$

where the relationship $A_\varphi = \zeta(r)\sin\theta$ was used. Substituting Eqs. (5.46) and (5.51) leads to

$$V_{22}^{(3)} = \frac{101\alpha(Z\alpha)^3}{55296}. \quad (5.55)$$

Upon combining Eqs. (5.49), (5.53) and (5.55) we are now in position to calculate $V_{22}$ of (5.48) and thereby to find the correction $\varepsilon_2$ to the energy of the $2p_{1/2}$ level by (5.28):

$$\varepsilon_2 = -\alpha(1+\zeta_\varphi)\frac{93}{512}Z\alpha + \alpha\left(\bar{c}_1 - \zeta_H\bar{a}_H - \zeta_\varphi\bar{c}_3 + \frac{101}{55296}\right)(Z\alpha)^3 + O(\alpha^6). \quad (5.56)$$

By analogy with Eq. (5.41), in the λ-renormalization of (5.6) for this state we put $\zeta_\varphi = -1$ and

$$\zeta_H = \frac{1}{\bar{a}_H}\left(\bar{c}_1 - \zeta_\varphi \bar{c}_3 + \frac{101}{55296}\right). \quad (5.57)$$

The second correction $\varepsilon_2^{(2)}$ to the energy renormalizes the constants $\zeta_\varphi$ and $\zeta_H$ similarly to the $2s_{1/2}$ level. Upon choosing the new constants $\zeta_\varphi$ and $\zeta_H$ appropriately, in lieu of (5.45) we shall have

$$\varepsilon_1 = -\alpha\bar{a}_H\zeta_1(Z\alpha)^4. \quad (5.58)$$

If we compare Eq. (5.58) with results of Ref. [7] as above, we shall obtain that $\zeta_1 \approx 0.3$ for the $2p_{1/2}$ level.

## 6. Concluding remarks

The factual development of SQED began in the late 1940s when the anomalous correction to the electron magnetic moment and the Lamb shift were calculated and the calculations were confirmed by experiment. In the present paper we have demonstrated how the anomalous electron magnetic moment and the Lamb shift can be obtained in NAQED in a first approximation. In contradistinction to SQED we have not met in our calculations with divergent integrals characteristic of SQED. Subsequent approximations must involve the bispinor $\psi_2$ and four-vector $v_\mu$ that figure in the basic equations of NAQED [1], which will essentially complicate the calculations. It is worth remarking that in SQED too the subsequent approximations demand very lengthy calculations.

SQED enables one to compute a limited number of quantities that can be observed in experiment because methods of SQED are based on the use of $q$-numbers whereas experimental data can be expressed in terms of $c$-numbers alone. Therefore, solely results that can, in SQED, be expressed in terms of *c*-numbers may be verified in experiment. For example, as mentioned in Introduction, although SQED indicates that the electron is not point-like, its methods give no way of determining the electron structure and its dimensions. At the same time results already obtained in SQED may supplement calculations carried out in NAQED, which is seen from Eqs. (5.45) and (5.58).

It is shown in Ref. [9] that the magnetic moment of the electron is not unique and depends on the external field in which the electron is placed. This is due to the fact that the field deforms the electron, which provokes a change of the magnetic moment. From this point of view the fact proven in Sec. 3 that the normal magnetic moment of a free electron and the normal magnetic moment of the same electron placed in a magnetic field are identical is not trivial.

Paper I is devoted to the electron and other leptons. An analogous approach is developed as to the proton and other hadrons in Ref. [10]. A characteristic feature of the proton is the fact that its magnetic moment differs substantially from the nuclear magneton whereas the electron magnetic moment is approximately equal to the Bohr magneton. For this reason the Dirac equation for the proton is supplemented in [10] by an extra term which yields an additional magnetic moment. Figuratively speaking, the whole of the proton magnetic moment is anomalous. This essentially affects properties of the proton as compared with the ones of the electron.

**Appendix A**

The scalar potential $\varphi$ can be computed upon solving the Poisson equation

$$\nabla^2 \varphi = -4\pi\rho, \tag{A.1}$$

where $\rho = |\psi|^2$ is the density of charge in our dimensionless units. In the non-relativistic limit we have from (5.11) that $\rho = f_1^2 / 4\pi$, so that Eq. (A.1) becomes

$$\nabla^2 \varphi = -f_1^2 \tag{A.2}$$

with $f_1$ from (5.30). Since $f_1$ does depend upon angles, Eq. (A.1) transforms into an ordinary differential equation, namely,

$$\frac{1}{r^2} \frac{d}{dr}\left(r^2 \frac{d\varphi}{dr}\right) = -f_1^2(r). \tag{A.3}$$

It follows from this that

$$\frac{d\varphi}{dr} = -\frac{1}{r^2} \int^r x^2 f_1^2(x) dx. \tag{A.4}$$

Calculating the integral with $f_1$ from (5.30) one finally arrives at the scalar potential in the state $2s_{1/2}$:

$$\varphi(r) = \frac{1}{r} - \frac{1}{r}\left(1 + \frac{3}{4}Z\alpha r + \frac{1}{4}(Z\alpha r)^2 + \frac{1}{8}(Z\alpha r)^3\right)e^{-Z\alpha r}. \tag{A.5}$$

Here the constants of integration are so chosen that $\varphi(0)$ is finite and $\varphi(\infty) = 0$.

In order to find the scalar potential created by the electron in the $2p_{1/2}$ state one should replace $f_1$ in (A.4) by $f_2$ of (5.46) with the result

$$\varphi(r) = \frac{1}{r} - \frac{1}{r}\left(1 + \frac{3}{4}Z\alpha r + \frac{1}{4}(Z\alpha r)^2 + \frac{1}{24}(Z\alpha r)^3\right)e^{-Z\alpha r}. \tag{A.6}$$